\documentclass{article}
\usepackage{amssymb}
\begin{document}
\title{Active versus passive transformations in the presence of a magnetic field}

\author{G.F.\ Torres del Castillo \\ Departamento de F\'isica Matem\'atica, Instituto de Ciencias \\
Universidad Aut\'onoma de Puebla, 72570 Puebla, Pue., M\'exico \\[2ex]
D.A.\ Rosete~\'Alvarez \\ Facultad de Ciencias F\'isico Matem\'aticas \\ Universidad Aut\'onoma de Puebla, 72570 Puebla, Pue., M\'exico}

\maketitle

\begin{abstract}
It is shown that, when there is a magnetic field present, in the framework of classical or quantum mechanics, the active translations differ from the passive ones and that the canonical momentum is not the generator of them. It is also shown that an infinitesimal generator of passive translations or rotations exists only if the magnetic field is invariant under these transformations.
\end{abstract}


\section{Introduction}
A recurrent topic in the textbooks on quantum mechanics is the representation of translations and rotations. Owing to its special features and many applications, the group of rotations receives particular attention, treating the group of translations in passing, many times as an introductory example of group representations.

In the textbooks on quantum mechanics it is customary to define the operator that represents a translation through ${\bf a}$ by
\begin{equation}
T({\bf a}) = \exp \left( - \frac{{\rm i}}{\hbar} {\bf a} \cdot {\bf p} \right), \label{to}
\end{equation}
where ${\bf p}$ is the linear momentum operator (see, e.g., Refs.\ \cite{GY,JJ,SW}). Furthermore, the commutativity of the spatial translations is employed to conclude that the Cartesian components of ${\bf p}$ must commute with each other. However, when there is a magnetic field present, the canonical momentum is a gauge-dependent quantity, so that, for each choice of the gauge, one obtains a different operator (\ref{to}). In fact, for a particle of mass $m$ and electric charge $e$, the canonical momentum, ${\bf p}$, is related to the so-called kinematical momentum
\begin{equation}
\mbox{\boldmath $\pi$} \equiv m \frac{{\rm d} {\bf r}}{{\rm d} t} \label{km}
\end{equation}
(which is gauge-independent) by
\begin{equation}
{\bf p} = \mbox{\boldmath $\pi$} + \frac{e}{c} {\bf A}, \label{cm}
\end{equation}
where ${\bf A}$ is a vector potential for the magnetic field present, ${\bf B} = \nabla \times {\bf A}$.

As we shall show, in the presence of a magnetic field, one has to distinguish between active and passive translations, and none of the infinitely many canonical momenta is the infinitesimal generator of them. In spite of its geometrical nature, in previous works the choice of the generator of translations has been based on algebraic arguments (see, e.g., Refs.\ \cite{Br,Za,Ja}).

Throughout this paper we make use of Cartesian coordinates and Cartesian components of vectors only.

\section{Translations in a uniform magnetic field in the framework of classical mechanics}
It will be very instructive to begin by studying the translations when there is a magnetic field present in the context of classical mechanics, using the Hamiltonian formalism. As pointed out above, the canonical momentum is gauge-dependent. Under a gauge transformation, ${\bf A} \mapsto {\bf A} + \nabla \xi$, where $\xi$ is an arbitrary differentiable function of the coordinates only, the Cartesian components of the canonical momentum transform according to $p_{i} \mapsto p_{i} + \partial (e \xi/c)/\partial x_{i}$ [see Eq.\ (\ref{cm})], but this corresponds to a canonical transformation since, if we denote the Poisson bracket by $\{ \, , \, \}$,
\begin{eqnarray*}
\{ p_{i} + \partial (e \xi/c)/\partial x_{i}, p_{j} + \partial (e \xi/c)/\partial x_{j} \} & = & \{ p_{i}, \partial (e \xi/c)/\partial x_{j} \} + \{ \partial (e \xi/c)/\partial x_{i}, p_{j} \} \\
& = & - \frac{\partial^{2} (e \xi/c)}{\partial x_{i} \partial x_{j}} + \frac{\partial^{2} (e \xi/c)}{\partial x_{j} \partial x_{i}} = 0
\end{eqnarray*}
and
\[
\{ x_{i}, p_{j} + \partial (e \xi/c)/\partial x_{j} \} = \{ x_{i}, p_{j} \} = \delta_{ij}.
\]

Since the Poisson bracket is invariant under canonical transformations, it follows that the Poisson bracket is gauge-invariant. In other words, the Poisson bracket of any two functions does not depend on the choice of the vector potential, which enters in the definition of the canonical momenta.

Any differentiable function, $G(x_{i}, p_{i})$, generates a (possibly local) one-parameter group of canonical transformations defined by
\begin{equation}
\frac{{\rm d} x_{i}(s)}{{\rm d} s} = \frac{\partial G}{\partial p_{i}}, \qquad \frac{{\rm d} p_{i}(s)}{{\rm d} s} = - \frac{\partial G}{\partial x_{i}}, \label{genf}
\end{equation}
with the initial conditions $x_{i}(0) = x_{i}$, $p_{i}(0) = p_{i}$. Equivalently, we can write Eqs.\ (\ref{genf}) in the form
\begin{equation}
\frac{{\rm d} x_{i}(s)}{{\rm d} s} = \{ x_{i}, G \}, \qquad \frac{{\rm d} p_{i}(s)}{{\rm d} s} = \{ p_{i}, G \}.
\label{genpb}
\end{equation}
Now, for translations along the $x_{k}$-axis we must have
\[
x_{i}(s) = x_{i} + s \, \delta_{ik},
\]
so that the parameter $s$ measures the length of the displacement. According to the first equation in (\ref{genf}), this implies that the generator of such translations must be given by
\begin{equation}
G_{k} = p_{k} + f(x_{1}, x_{2}, x_{3}), \label{gtr}
\end{equation}
where $f$ is some function of three variables. In order to find the expression for $f$, {\em it is necessary to specify the effect of the translations on the momentum of the particle.}

\subsection{Passive translations}
In the case of passive translations (the particle is observed from two different frames, one of which is translated with respect to the other) we expect that the Cartesian components of the {\em kinematical}\/ momentum be invariant, i.e.,
\begin{equation}
\{ \pi_{i}, G_{k} \} = 0 \label{pasg}
\end{equation}
($i = 1, 2, 3$), which, by virtue of (\ref{cm}) and (\ref{gtr}), amounts to $\{ p_{i} - (e/c) A_{i}, p_{k} + f \} = 0$, that is,
\begin{equation}
- \frac{e}{c} \frac{\partial A_{i}}{\partial x_{k}} - \frac{\partial f}{\partial x_{i}} = 0. \label{efe}
\end{equation}
(Note that the presence of the vector potential in this equation implies that $f$ is gauge-dependent; but this is precisely what is to be expected, since the first term on the right-hand side of (\ref{gtr}) is also gauge-dependent, and in this manner $G_{k}$ turns out to be gauge-independent.) From Eq.\ (\ref{efe}) it follows that
\[
\frac{\partial^{2} f}{\partial x_{j} \partial x_{i}} = - \frac{e}{c} \frac{\partial^{2} A_{i}}{\partial x_{j} \partial x_{k}}
\]
and, therefore, the equality of the mixed second partial derivatives of $f$ amounts to the condition $\partial B_{i}/\partial x_{k} = 0$, for $i = 1, 2, 3$; that is, the magnetic field must be invariant under the translations along the $x_{k}$-axis \cite{CC}. If this condition is not satisfied, then Eq.\ (\ref{efe}) has no solution, which means that these passive translations are not canonical transformations (see also the discussion below).

In particular, for a {\em uniform}\/ field, Eq.\ (\ref{efe}) has solution for all values of $k$, and if we choose ${\bf A} = \frac{1}{2} {\bf B} \times {\bf r}$, then the solution of (\ref{efe}) is given by $f = (e/2c) \, \varepsilon_{klm} B_{l} x_{m}$, and the generating function of the passive translations along the $x_{k}$-axis is
\begin{equation}
G_{k} = p_{k} + \frac{e}{2c} \varepsilon_{klm} B_{l} x_{m} = \pi_{k} + \frac{e}{c} \varepsilon_{klm} B_{l} x_{m}. \label{gpk}
\end{equation}
(Note that $G_{k}$ and $p_{k}$ are not related by a gauge transformation.)

It is important to point out that, in spite of the fact that the translations must commute, the Poisson brackets between the functions $G_{k}$ are different from zero, viz.,
\begin{equation}
\{ G_{i}, G_{j} \} = \{ p_{i} + \frac{e}{2c} \varepsilon_{ilm} B_{l} x_{m}, p_{j} + \frac{e}{2c} \varepsilon_{jkn} B_{k} x_{n} \} = - \frac{e}{c} \varepsilon_{ijl} B_{l}. \label{crg}
\end{equation}
However, the fact that these brackets are constant implies that the transformations generated by these functions (on the phase space) do commute.

If the particle only interacts with the magnetic field, its Hamiltonian is given by
\begin{equation}
H = \frac{1}{2m} \left( {\bf p} - \frac{e}{c} {\bf A} \right)^{2} = \frac{1}{2m} \mbox{\boldmath $\pi$}^{2} \label{ham}
\end{equation}
and from Eq.\ (\ref{pasg}) we see that $\{ G_{k}, H \} = 0$, i.e., $G_{k}$ is conserved if the magnetic field is invariant under the translations along the $x_{k}$-axis. (Note that, in a naive way, one would expect a relationship between the translational symmetry of the mechanical system and the conservation of the linear momentum; as we have shown, this symmetry is related with the conservation of the generators of passive translations, which differ from the canonical momenta. Cf.\ also Refs.\ \cite{Br,Za,GM}.)

A finite passive translation along the $x_{k}$-axis is given by
\[
x_{i}(s) = x_{i} + s \, \delta_{ik}, \qquad \pi_{i}(s) = \pi_ {i}
\]
($i = 1, 2, 3$) hence,
\[
p_{i}(s) = p_{i} + \frac{e}{c} [A_{i}(x_{j}(s)) - A_{i}(x_{j})]
\]
and, therefore, $\{ x_{i}(s), p_{j}(s) \} = \delta_{ij}$, but
\[
\{ p_{i}(s), p_{j}(s) \} = \frac{e}{c} \varepsilon_{ijl} [B_{l}(x_{m}) - B_{l}(x_{m}(s))],
\]
which is equal to zero if and only if the Cartesian components of ${\bf B}$ do not depend on $x_{k}$. Thus, a passive translation is a canonical transformation if and only if the magnetic field is invariant under the translation.

\subsection{Active translations}
When a charged particle moves in a magnetic field, there appears a magnetic force given by
\[
\frac{{\rm d} \pi_{i}}{{\rm d} t} = \frac{e}{c} \varepsilon_{ilm} \frac{{\rm d} x_{l}}{{\rm d} t} B_{m}.
\]
That is, a change in the coordinates of the particle (with respect to a single frame of reference), produces a change in its kinematical momentum,
\begin{equation}
{\rm d} \pi_{i} = \frac{e}{c} \varepsilon_{ilm} B_{m} {\rm d} x_{l}. \label{lor}
\end{equation}
Then, by combining Eqs.\ (\ref{cm}), (\ref{genpb}), (\ref{gtr}), and (\ref{lor}), one finds that, without having to fix the gauge,
\[
f = - \frac{e}{c} A_{k}.
\]
(Note that, by contrast with the case of the passive translations considered in the previous subsection, in this instance there are no integrability conditions on the magnetic field.) Therefore, substituting into (\ref{gtr}), one concludes that
\begin{equation}
G_{k} = p_{k} - \frac{e}{c} A_{k} = \pi_{k} \label{gtra}
\end{equation}
is the generating function of active translations along the $x_{k}$-axis. As in the case of (\ref{gpk}), the last expression is gauge-independent. Also in this case, $G_{k}$ and $p_{k}$ are not related by a gauge transformation.

Again, even though the spatial translations commute, the Poisson brackets between the components $\pi_{i}$ are different from zero,
\begin{equation}
\{ \pi_{i}, \pi_{j} \} = \{ p_{i} - \frac{e}{c} A_{i}, p_{j} - \frac{e}{c} A_{j} \} = \frac{e}{c} \varepsilon_{ijl} B_{l}. \label{crga}
\end{equation}
Only if the field is uniform these brackets are constant, and only in that case the transformations on the phase space generated by the $\pi_{i}$ do commute.

To end this section, we point out that, for any choice of the vector potential, the Cartesian components of the canonical momentum satisfy the relations
\[
\{ p_{i}, p_{k} \} = 0,
\]
which, together with $\{ x_{i}, p_{k} \} = \delta_{ik}$, means that under the translations generated by $p_{k}$ the three components of ${\bf p}$ do not change. However, these translations do not seem interesting from the geometrical point of view, since ${\bf p}$ is gauge-dependent.

\section{Translations in a uniform magnetic field in the framework of quantum mechanics}
Now we shall repeat the steps followed in the foregoing section, studying the translations in the context of quantum mechanics, when there is a magnetic field present.

\subsection{Passive translations in quantum mechanics}
Guided by the results of the preceding section, we shall say that the Hermitean operator $G_{k}$ is the generator of passive translations along the $x_{k}$-axis if
\begin{equation}
[x_{i}, G_{k}] = {\rm i} \hbar \delta_{ik}, \qquad [\pi_{i}, G_{k} ] = 0, \label{gpt}
\end{equation}
for $i = 1, 2, 3$, with $\pi_{i} = p_{i} - \frac{e}{c} A_{i}$ (see Eqs.\ (\ref{cm}) and (\ref{pasg})). Then, making use of the Jacobi identity
\[
[\pi_{i}, [\pi_{j}, G_{k}]] + [\pi_{j}, [G_{k}, \pi_{i}]] + [G_{k}, [\pi_{i}, \pi_{j}]] = 0
\]
and the fact that
\begin{equation}
[\pi_{i}, \pi_{j}] = {\rm i} \hbar \frac{e}{c} \varepsilon_{ijl} B_{l} \label{crpi}
\end{equation}
(cf.\ Eq.\ (\ref{crga})), we obtain $[G_{k}, B_{l}] = 0$, that is, $\partial B_{l}/\partial x_{k} = 0$, for $l = 1, 2, 3$, which means that the existence of $G_{k}$ implies that the magnetic field must be invariant under the translations along the $x_{k}$-axis.

Thus, in the particular case of a uniform magnetic field, Eqs.\ (\ref{gpt}) are integrable for $k = 1, 2, 3$, and one can verify that their solution is given by the gauge-independent expression
\[
G_{k} = \pi_{k} + \frac{e}{c} \varepsilon_{klm} B_{l} x_{m}
\]
(cf.\ Eq.\ (\ref{gpk})). The commutators between the operators $G_{k}$ are multiples of the identity operator,
\begin{equation}
[G_{i}, G_{j}] = - {\rm i} \hbar \frac{e}{c} \varepsilon_{ijl} B_{l}, \label{crgq}
\end{equation}
and therefore the passive translation operators
\begin{equation}
T_{G}({\bf a}) \equiv \exp \left( - \frac{{\rm i}}{\hbar} {\bf a} \cdot {\bf G} \right), \label{pto}
\end{equation}
commute up to a phase factor. In fact, making use of the identity $\exp A  \exp B = \exp (A + B) \exp ({\textstyle \frac{1}{2}} [A, B])$, which is valid if $[A, B]$ commutes with $A$ and $B$, we obtain
\begin{equation}
T_{G}({\bf a}) \; T_{G}({\bf b}) = \exp \left( \frac{{\rm i} e}{2 \hbar c} {\bf a} \times {\bf b} \cdot {\bf B} \right) T_{G}({\bf a} + {\bf b}). \label{ray}
\end{equation}
The real number $\frac{1}{2} {\bf a} \times {\bf b} \cdot {\bf B}$ is the magnetic flux through the triangle with vertices at the origin and the points ${\bf a}$ and ${\bf a + b}$.

The translation operators (\ref{pto}) have been employed in the study of electrons in a periodic potential (see, e.g., Refs.\ \cite{Br,Za} and the references cited therein), without recognizing its geometrical meaning, looking for operators that commute with the Hamiltonian. In fact, in Ref.\ \cite{Za} it is claimed that in the presence of a uniform magnetic field, the Hamiltonian is not invariant under the translation group. As we have shown, by virtue of the conditions $[\pi_{i}, G_{k}] = 0$, the Hamiltonian is invariant under the passive translation operators (\ref{pto}), which, owing to (\ref{ray}), form a ray representation of the translation group. On the other hand, in Ref.\ \cite{Br} it is pointed out that the motion of a charged particle in a magnetic field produces a change in the kinematical momentum of the particle (see Eq.\ (\ref{lor})), without realizing that the translations considered there leave the kinematical momentum invariant.

\subsection{Active translations in quantum mechanics}
As in the case of classical mechanics, the kinematical momentum, \mbox{\boldmath $\pi$}, will be the infinitesimal generator of active translations (and is defined even if the field is nonuniform). As a consequence of the commutation relations (\ref{crpi}), the active translation operators
\[
T_{\pi}({\bf a}) \equiv \exp \left( - \frac{{\rm i}}{\hbar} {\bf a} \cdot \mbox{\boldmath $\pi$} \right)
\]
do not commute with each other

The kinematical momentum as a generator of translations was employed by Jackiw \cite{Ja} in his derivation of the Dirac quantization condition. Jackiw showed that in the field of a magnetic monopole, the translations generated by \mbox{\boldmath $\pi$} satisfy the associative law (without extra phase factors) if the charge of the monopole satisfies the Dirac quantization condition.

\section{Passive rotations in the presence of a magnetic field}
According to the discussion above, we can define the generator, $L_{k}$, of {\em passive}\/ rotations about the $x_{k}$-axis by the conditions
\begin{equation}
\{ x_{i}, L_{k} \} = \varepsilon_{ikl} x_{l}, \qquad \{  \pi_{i}, L_{k} \} = \varepsilon_{ikl} \pi_{l}, \label{rc}
\end{equation}
in classical mechanics, and by
\begin{equation}
[x_{i}, L_{k}] = {\rm i} \hbar \varepsilon_{ikl} x_{l}, \qquad [\pi_{i}, L_{k}] = {\rm i} \hbar \varepsilon_{ikl} \pi_{l}, \label{rq}
\end{equation}
in quantum mechanics.

From the first equation in (\ref{rc}) it follows that
\[
L_{k} = \varepsilon_{kij} x_{i} p_{j} + g(x_{1}, x_{2}, x_{3}),
\]
where $g$ is a real-valued function of three variables. Substituting this expression into the second equation in (\ref{rc}) we obtain the conditions
\begin{equation}
\frac{\partial g}{\partial x_{i}} = \frac{e}{c} \left( \varepsilon_{ikl} A_{l} - \varepsilon_{klj} x_{l} \frac{\partial A_{i}}{\partial x_{j}} \right). \label{rcon}
\end{equation}
The equality of the mixed second partial derivatives of $g$ imply that ${\bf B}$ must be invariant under the rotations about the $x_{k}$-axis. If the magnetic field is not invariant under these rotations, then the passive rotations about the $x_{k}$-axis are not canonical transformations.

The expression inside the parenthesis in Eq.\ (\ref{rcon}) is equal to zero if and only if the vector potential {\bf A} is invariant under the rotations about the $x_{k}$-axis and, therefore, in those cases, we can take $g = 0$. For example, a uniform magnetic field ${\bf B} = B {\bf k}$, parallel to the $x_{3}$-axis, is invariant under the rotations about the $x_{3}$-axis and, choosing ${\bf A} = \frac{1}{2} {\bf B} \times {\bf r}$, the vector potential itself is invariant under these rotations, so that $g$ can be chosen equal to zero. Then,
\begin{equation}
L_{3} = x_{1} p_{2} - x_{2} p_{1} = x_{1} \pi_{2} - x_{2} \pi_{1} + \frac{eB}{2c} (x_{1}{}^{2} + x_{2}{}^{2}), \label{l3}
\end{equation}
but $L_{1}$ and $L_{2}$ do not exist. The last expression in (\ref{l3}) is gauge-independent.

Another axially symmetric magnetic field is that produced by a magnetic dipole. The vector potential for a point magnetic dipole at the origin, with dipole moment \mbox{\boldmath $\mu$}, can be chosen as ${\bf A} = \mbox{\boldmath $\mu$} \times {\bf r}/r^{3}$, which is also invariant under the rotations about \mbox{\boldmath $\mu$}.

If the Hamiltonian is given by Eq.\ (\ref{ham}), then, by virtue of (\ref{rc}) or (\ref{rq}), $L_{k}$ is a constant of motion if the magnetic field is invariant under rotations about the $x_{k}$-axis.

\section{Concluding remarks}
As we have shown, in the Hamiltonian formulation of classical mechanics and in quantum mechanics, the definition of a translation or of a rotation requires the specification of its effect on the kinematical momentum.

Another important conclusion is that one cannot impose the commutativity of the operators representing translations based on the fact that the translations in Euclidean space do commute.

\section*{Acknowledgement}
One of the authors (D.A.R.A.) wishes to thank the Consejo Nacional de Ciencia y Tecnolog\'ia for financial support.

\end{document}